\pdfoutput=1

\documentclass[11pt]{article}

\usepackage[preprint]{acl}

\usepackage{times}
\usepackage{latexsym}
\usepackage{amsmath}
\newtheorem{lemma}{Lemma}
\newtheorem{proposition}{Proposition}
\newtheorem{theorem}{Theorem}
\usepackage{amsfonts}
\usepackage[T1]{fontenc}

\usepackage[utf8]{inputenc}

\usepackage{microtype}

\usepackage{inconsolata}

\usepackage{graphicx}

\usepackage{booktabs} 
\usepackage{subcaption}

\title{Heterogeneity-Aware Dataset Scheduling for Efficient Audio Large Language Model Training}

\author{
  Yanru~Wu\textsuperscript{1}\hspace{0.5cm}
  Jianning~Wang\textsuperscript{2}\hspace{0.5cm}
  Chongxin~Gan\textsuperscript{3}\hspace{0.5cm}
  Yang~Li\textsuperscript{1}\thanks{Corresponding Author}\\[10pt]
  \textsuperscript{1}\textit{Shenzhen International Graduate School, Tsinghua University}\\
  \textsuperscript{2}\textit{Independent Researcher}\\
  \textsuperscript{3}\textit{The Hong Kong Polytechnic University}
}

\begin{document}
\maketitle
\begin{abstract}
Training general-purpose Audio Large Language Models (ALLMs) across diverse datasets is essential for holistic audio understanding, yet it faces significant challenges due to dataset heterogeneity, which often leads to conflicting gradients and slow convergence. Despite its impact, how to explicitly manage this heterogeneity during training remains underexplored, with current practices relying primarily on uniform mixture. In this work, we analyze multi-dataset AudioQA training from a convergence perspective and propose Grouped Sequential Training (GST). GST strategically organizes datasets into affinity-aware groups and introduces them via a progressive scheduling protocol, effectively balancing the stability of parallel training with the efficiency of sequential optimization. To ensure scalability, we develop gradient-based affinity metrics that capture inter-dataset relationships without the prohibitive cost of empirical transferability estimation. Extensive evaluations on 14 AudioQA datasets spanning speech, music, and environmental sounds demonstrate that GST achieves 30--40\% faster convergence than standard parallel training while maintaining or even surpassing the performance of mix-all training. Our results provide both theoretical insights and a practical, model-agnostic framework for efficient large-scale ALLM optimization.

\end{abstract}

\section{Introduction}

Large language models (LLMs) have recently been extended beyond text to incorporate audio perception, enabling progress in tasks such as classification, captioning, and question answering \cite{kong2024audio, hurst2024gpt}. Among these, AudioQA \cite{openAQA} has attracted increasing attention as a unified data format for training Audio LLMs (ALLMs), consolidating diverse audio tasks into a single framework and supporting broad application scenarios from assistive technology to human–computer interaction. While recent efforts have integrated pretrained audio encoders into LLMs to build general-purpose AudioQA models \cite{chu2024qwen,tang2024salmonn}, scaling across datasets with disparate domains, annotation styles, and difficulty levels introduces significant yet often overlooked training inefficiencies.

Multi-task learning \cite{zhang2021survey} provides a natural theoretical lens for generalization across diverse datasets, where in practice each dataset is often treated as an instantiation of a task. 
Extensive explorations have been conducted in traditional machine learning \cite{standley2020tasks, liu2019end}, encompassing investigations of curriculum learning \cite{soviany2022curriculum}, task sampling \cite{sharma2017learning}, and various training paradigms \cite{elich2024examining}. However, these approaches typically rely on task-specific assumptions, architectural modifications, or auxiliary training mechanisms, limiting their applicability to general ALLMs.\footnote{See Sec.~\ref{sec:rel-training} for further discussion.}
Consequently, most existing ALLM training pipelines resort to naive strategies such as uniform dataset mixing, and fail to account for substantial data heterogeneity.
Although such mixing ensures task coverage, it often induces suboptimal training dynamics characterized by redundant updates and slow convergence. These inefficiencies are further exacerbated in ALLMs by the inherent complexities of long audio sequences, cross-modal alignment, and the sheer scale of the models.

Recent advances in taskonomy and transferability estimation \cite{zamir2018taskonomy, wang2025understanding} have established that dataset-level relationships can be successfully quantified. Yet despite this progress, a systematic understanding of how dataset-level heterogeneity governs the training of ALLMs remains elusive. Current analyses are largely restricted to post-hoc evaluations, leaving a gap in understanding how these heterogeneous traits directly impact real-time training dynamics, and more importantly, how they can inform the design of training strategies. Building on this intuition, this work addresses a central question: How can these dataset properties be explicitly leveraged to schedule efficient multi-dataset training for ALLMs, thereby accelerating convergence without architectural overheads?

To address this, we investigate multi-dataset training for ALLMs through the lens of convergence analysis. We begin by theoretically analyzing the convergence behaviors of different training regimes, focusing on how gradient variance and dataset heterogeneity affect training dynamics. Our analysis reveals the relative advantages of parallel versus sequential training under differing heterogeneity degrees, which then motivates our proposal of Grouped Sequential Training (GST). As a paradigm that interpolates between parallel and sequential optimization, GST organizes datasets into groups with reduced intra-group heterogeneity. By executing sequential updates at the group level, this strategy yields provably tighter convergence bounds in realistic multi-dataset settings.
For practical implementation, we cluster datasets using metrics that capture optimization-level disparities, such as gradient-based distance and empirical transferability. To further scale the framework to large-scale ALLMs, we introduce an efficient progressive training procedure that approximates grouped sequential optimization while minimizing computational overhead and mitigating catastrophic forgetting. Importantly, our approach is model-agnostic and requires no architectural modifications. Extensive evaluations across 14 AudioQA datasets demonstrate that our method significantly accelerates convergence compared to naive multi-task baselines, while simultaneously maintaining or enhancing performance across diverse domains. Our contributions can be summarized as follows:

\begin{itemize}
    \item \textbf{Theoretical Framework:} We establish a convergence analysis for ALLMs that rigorously characterizes the impact of gradient variance and dataset heterogeneity.
    \item \textbf{Paradigm Shift:} We propose Grouped Sequential Training (GST), which interpolates between parallel and sequential optimization and moves dataset affinity estimation from post-hoc analysis into training loop design.
    \item \textbf{Scalable Implementation:} We introduce a progressive training procedure that approximates the optimal grouping strategy, enabling efficient large-scale training without architectural changes.
    \item \textbf{Benchmark \& Insights:} We consolidate 14 datasets into a unified AudioQA format, providing systematic insights and actionable strategies for heterogeneous multi-dataset training.
\end{itemize}

\begin{figure*}[htbp]
\centering

\includegraphics[width=0.8\textwidth]{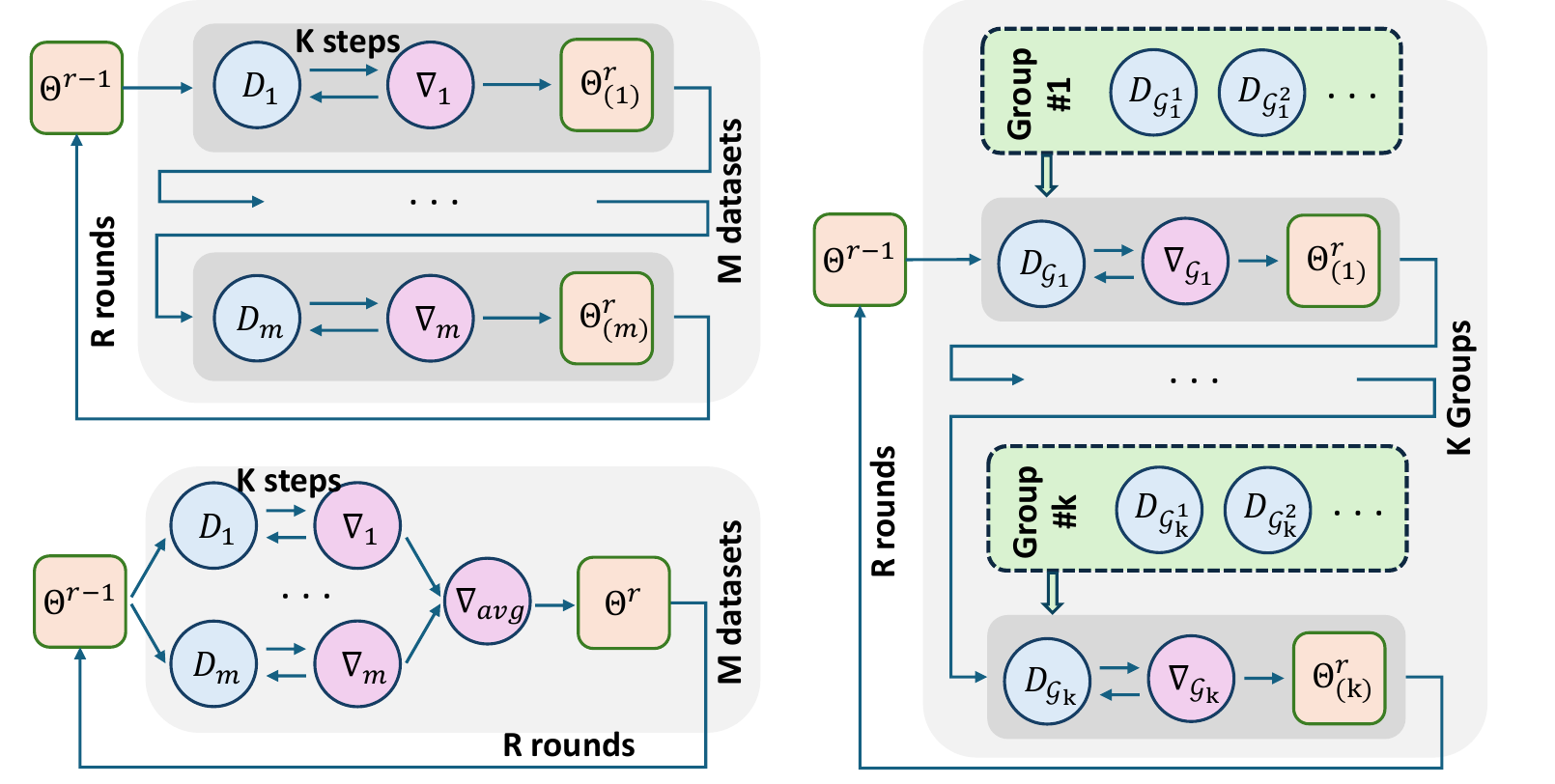}

\caption{\textbf{Illustration of sequential (left-up), parallel (left-down), and grouped sequential training (right).}}
\label{fig:framework}
\end{figure*}

\section{Related Works}

\subsection{Audio Large Language Model}

ALLMs have transformed audio-related tasks by demonstrating remarkable zero-shot performance and classification, understanding, and reasoning capabilities. Existing approaches typically adopt either an encoder-only or an encoder-decoder architecture \cite{saeed2021contrastive}. The encoder-only paradigm employs a dual-tower framework where the text and audio encoders process paired text-audio data. Following the training objective of Contrastive Language-Image Pre-training (CLIP) \cite{radford2021learning}, such models are trained to learn a joint embedding space by drawing the embeddings of paired data closer while pushing apart those of unpaired data. 

In parallel, driven by the success of vision-language models, various encoder-decoder ALLMs have been proposed to leverage the large language model for audio comprehension. Prior works such as SALMONN \cite{tang2024salmonn}, Pengi \cite{deshmukh2023pengi}, Qwen-Audio \cite{chu2023qwen, chu2024qwen}, and the Audio Flamingo series \cite{kong2024audio, ghosh2025audio, goel2025audio} have achieved state-of-the-art performances across a wide range of audio tasks. Those methods integrate an audio encoder with a decoder-only language model in a cascade manner. Unlike encoder-only methods that solely rely on contrastive training objectives, encoder–decoder ALLMs are trained with generative objectives, such as masked reconstruction or next-token prediction. This paradigm enables the model to capture richer semantic audio information, facilitating more flexible reasoning and instruction-following abilities.

\subsection{Training Strategies for Multi-Dataset Learning}
\label{sec:rel-training}

Training models on multiple datasets is a fundamental challenge across multi-task \cite{zhang2021survey}, continual \cite{kirkpatrick2017overcoming}, and federated learning \cite{li2020federated}. In the context of LLMs, research has increasingly shifted from purely architectural innovations \cite{houlsby2019parameter, shazeer2017outrageously} to data-centric optimization strategies. While substantial progress has been made in data selection and sampling weighting \cite{xie2023doremi, chen2023skill, li2024datacomp} to curate optimal training mixtures, the role of learning scheduling remains underexplored. 

Existing paradigms generally fall into two categories: parallel and sequential training (Fig.~\ref{fig:framework}). Parallel training jointly optimizes all datasets via uniform mixing or weighted sampling \cite{raffel2020exploring}. While stable, it often suffers from gradient interference \cite{yu2020gradient}, where conflicting optimization directions from diverse domains lead to suboptimal convergence. In contrast, sequential training introduces datasets chronologically \cite{bengio2009curriculum}. While this can prioritize high-quality data to bootstrap representations, it risks catastrophic forgetting of earlier tasks and often lacks a principled basis for determining the optimal sequence. Most theoretical analyses \cite{koloskova2020unified, li2023convergence}, often established within the context of federated optimization, are confined to idealized formulations, focusing on the two extremes: fully parallel or strictly sequential regimes. Strategies that interpolate between these extremes, such as data-driven grouping \cite{standley2020tasks} or stage-wise scheduling \cite{zamir2018taskonomy}, often lack systematic, convergence-oriented analyses tailored for dataset-level ALLM training. Furthermore, their post-hoc design and task dependency hinder their integration into the dynamic training loops of large-scale models.

\section{Problem Setup}

\subsection{Formulation}

Modern ALLMs unify diverse audio understanding tasks through a shared language interface, enabling natural language reasoning
over acoustic signals \cite{latif2023sparks}. Consider the multi-dataset training problem for AudioQA. Let $\mathcal{D}=\{D_m\}_{m=1}^{M}$ denote a collection of $M$ datasets, where each dataset $D_m = \{(a_i, x_i, y_i)\}_{i=1}^{N_m}$ consists of audio inputs $a_i$ and their corresponding textual question-answer pairs $(x_i, y_i)$.

Given a model $f$ parameterized by $\Theta$, our objective is to minimize the aggregate loss across all tasks:
\begin{align}
    \min_{\Theta}\left\{F(\Theta):=\frac{1}{M}\sum_{m=1}^M F_m(\Theta)\right\},
\end{align}
\noindent where $F_m(\theta)$ denotes the task-specific loss on $D_m$, measured by cross-entropy in our ALLM training. Following standard ALLM architectures \cite{tang2024salmonn}, we initialize $f$ from a pretrained LLM backbone (e.g., a LLaMA-style decoder) and extend it to the audio modality with audio feature extractors and learnable adapters (e.g., LoRA layers), while the backbone weights maintain frozen.

\subsection{Characterizing Dataset Heterogeneity}

Despite sharing a common input-output format, AudioQA datasets exhibit pronounced heterogeneity due to acoustic characteristics, annotation styles, and task complexities. To rigorously quantify this, we adopt the gradient decomposition framework from federated optimization \cite{li2023convergence}, characterizing heterogeneity through gradient inconsistency across datasets:
\begin{align}
\nonumber
&\frac{1}{M}\sum_{m=1}^M\ 
\|\nabla F_m(\Theta) - \nabla F(\Theta)\|^2\\
&\le
\beta^2 \|\nabla F(\Theta)\|^2 + \zeta^2,
\label{eqn:gvariance}
\end{align}
\noindent where the constants $\beta$ and $\zeta$ capture the degree of dataset-level divergence. Intuitively, larger values of $\beta^2$ and $\zeta^2$ indicate stronger disagreement among gradients derived from different datasets, a phenomenon known to impede convergence in both sequential and parallel optimization regimes. In the context of ALLMs, such gradient inconsistency is not merely an artifact of distribution shifts but reflects intrinsic semantic disparities. For example, datasets emphasizing temporal localization may induce gradients that prioritize fine-grained acoustic cues, whereas captioning datasets favor global semantic representations. These differences persist even when datasets are projected onto a unified language output space, necessitating a more sophisticated training strategy than naive mixing.

\section{Methodology}


\subsection{Sequential and Parallel Training under Dataset Heterogeneity}
\label{sec: seq and par}


To analyze the impact of dataset heterogeneity, we first establish a theoretical foundation, measuring convergence by the minimum expected gradient norm across $R$ training rounds: 
\begin{align}
    \mathcal{C}(R):=\min_{0 \le r \le R} \mathbb{E}\big[\|\nabla F(\Theta^{(r)})\|^2\big],
\end{align}
where the convergence rate is characterized by the decay of $\mathcal{C}(R)$ with respect to $R$. 
We then adapt the theoretical findings from federated learning \cite{li2023convergence} to the multi-dataset setting by treating each dataset as a distinct ``client", simplifying the complex distributed dynamics into two primary factors - stochastic gradient variance and dataset heterogeneity. Specifically, we have the following assumptions and results:

\noindent \textbf{AS1 (Smoothness):} Each dataset-specific objective $F_m$ is $L$-smooth.

\noindent \textbf{AS2 (Bounded Variance)}: The gradient computed on each dataset has a variance bounded by $\sigma^2$.

\noindent  \textbf{AS3 (Bounded Heterogeneity):} Dataset-level heterogeneity is bounded by constants $\beta$ and $\zeta$ as defined in Eqn.~\eqref{eqn:gvariance}.

\begin{lemma}[Par. and Seq. Training Convergence]
\label{lemma:convergence}
Under Assumptions AS1--AS3, the convergence rates for parallel and sequential training over $M$ datasets satisfy:
\begin{align}
\mathcal{C}_{par}&= \mathcal{O}\!\left(
\frac{A\sigma}{R^{\frac{1}{2}}}
+ \frac{B\sigma^{\frac{2}{3}}+C\zeta^{\frac{2}{3}}}{R^{\frac{2}{3}}}
+ \frac{D(1 + \beta^2)}{R}
\right),
\label{eq:parallel_bound} \\
\mathcal{C}_{seq}&= \mathcal{O}\!\left(
\frac{A\sigma}{R^{\frac{1}{2}}}
+ \frac{B\sigma^{\frac{2}{3}}+C\zeta^{\frac{2}{3}}}{M^{\frac{1}{3}} R^{\frac{2}{3}}}
+ \frac{D(1 + \frac{\beta^2}{M})}{R}
\right),
\label{eq:sequential_bound}
\end{align}
\noindent where $A, B, C, D$ are positive constants depending on the smoothness $L$ and optimization hyper-parameters.
\end{lemma}

Comparing the bounds in Lemma~\ref{lemma:convergence} reveals a fundamental trade-off. Parallel training benefits from joint sampling, which stabilizes optimization and reduces the effective $\sigma^2$. However, it is fully exposed to the global heterogeneity terms $\beta^2$ and $\zeta^{2/3}$. In contrast, sequential training suppresses these heterogeneity effects by factors of $1/M$ and $1/M^{1/3}$, effectively ``decoupling" the interference between divergent datasets at the cost of losing joint variance reduction. Consequently, while parallel mixing is robust for mild heterogeneity, the sequential regime becomes theoretically superior as dataset divergence increases - as is common in complex ALLM tasks.
This observation suggests that neither extreme is globally optimal, motivating a strategy that balances variance reduction with heterogeneity suppression.

\subsection{Grouped Sequential Training (GST)}

AudioQA datasets often exhibit structured heterogeneity. For instance, datasets involving music perception share closer optimization trajectories than those requiring speech comprehension. To capture this structure and achieve the ``best of both worlds" discussed in Sec.~\ref{sec: seq and par}, we propose to partition the $M$ datasets into $K$ affinity-based groups $\{\mathcal{G}_k\}_{k=1}^K$ and conduct Grouped Sequential Training (GST), a hybrid paradigm that leverages local variance reduction while maintaining global heterogeneity suppression (Fig.~\ref{fig:framework}). For each group, the group-level objective is defined as:
\begin{align}
    F^{(k)}(\Theta)
    :=\frac{1}{M_k}
    \sum_{m \in \mathcal{G}_k} F_m(\Theta),
\end{align}
where $M_k$ is the number of datasets in $\mathcal{G}_k$ and $F^{(k)}$ inherits $L$-smoothness from its constituent datasets. The global gradient
variance can thus be decomposed as follows:
\begin{proposition}[Heterogeneity Decomposition]
\label{prop:decomposition}
For any partition $\{\mathcal{G}_k\}_{k=1}^K$, the global gradient variance can be decomposed into intra-group and inter-group components:
\begin{align}
\small
&\underbrace{\frac{1}{M} \sum_{m=1}^M \|\nabla F_m - \nabla F\|^2}_{\text{Global Variance}} \\ = &\underbrace{\frac{1}{M} \sum_{k=1}^K \sum_{m \in \mathcal{G}_k} \|\nabla F_m - \nabla F^{(k)}\|^2}_{\text{Intra-group Heterogeneity}} \\+& \underbrace{\frac{1}{M} \sum_{k=1}^K M_k \|\nabla F^{(k)} - \nabla F\|^2}_{\text{Inter-group Heterogeneity}}.
\label{eqn:variance_decompose_simplified}
\end{align}
\end{proposition}

Following this decomposition, we characterize heterogeneity bound at two levels: intra-group $(\beta_k, \zeta_k)$ and inter-group $(\beta_g, \zeta_g)$. Analogous to Eqn.~\eqref{eqn:gvariance}, they satisfy:
\begin{align}
\frac{1}{M_k}
\sum_{m \in \mathcal{G}_k}
&\|\nabla F_m(\Theta) - \nabla F^{(k)}(\Theta)\|^2 \nonumber\\
\le \ \beta_k^2& \|\nabla F^{(k)}(\Theta)\|^2 + \zeta_k^2,
\label{eq:group_heterogeneity} \\
    \frac{1}{K}\sum_{k=1}^K&\|\nabla F^{(k)}(\Theta)-\nabla F(\Theta)\|^2 \nonumber\\
    \le \ \beta_g^2 &\|\nabla F(\Theta)\|^2 + \zeta_g^2.
    \label{eqn:groupvariance}
\end{align}

We establish the relationship between group-wise and global heterogeneity from Prop.~\ref{prop:decomposition}:
\begin{lemma}[Heterogeneity Bound]
\label{lemma:group_heterogeneity}
For a grouping strategy that clusters datasets with high affinities, the intra-group and inter-group heterogeneity constants satisfy:
\begin{align}
\text{(Intra-group):} \quad &\beta_k^2 \le \beta^2, \quad \zeta_k^2 \le \zeta^2, \\
\text{(Inter-group):} \quad &\beta_g^2 \le \frac{M\beta^2}{K M_{min}} - \Gamma_{\mathcal{G}}, \nonumber\\
\quad &\zeta_g^2 \le \frac{M\zeta^2}{K M_{min}} - \Delta_{\mathcal{G}},
\end{align}
where $M_{min}$ is the smallest group size, and $\Gamma_{\mathcal{G}}, \Delta_{\mathcal{G}} \ge 0$ are the residual terms representing the heterogeneity absorbed within groups via affinity-based alignment.
\end{lemma}
The proof is given in  Appendix \ref{proof:lemma2}.


Based on Lemma \ref{lemma:group_heterogeneity}, GST satisfies the three fundamental optimization assumptions: (i) $F^{(k)}$ inherits $L$-smoothness; (ii) the stochastic gradient variance within groups is bounded by $\sigma_k^2<\sigma^2$ guaranteed by the bound of intra-group heterogeneity $(\beta_k, \zeta_k)$;\footnote{Refer to Appendix \ref{proof:variance} for the proof.} and (iii) inter-group heterogeneity is bounded by $(\beta_g, \zeta_g)$. We then apply the convergence results at the group level and have:
\begin{theorem}[Convergence of GST] 
\label{theorem:gst_bound} 
Under Assumptions AS1--AS3, the convergence rate of Grouped Sequential Training satisfies: 
\begin{equation} \mathcal{C}_{gst} = \mathcal{O}\!\left( \frac{A\sigma_g}{R^{\frac{1}{2}}} + \frac{B\sigma_g^{\frac{2}{3}} + C\zeta_g^\frac{2}{3}}{K^{\frac{1}{3}} R^{\frac{2}{3}}} + \frac{D(1 + \frac{\beta_g^2}{K})}{R} \right). \label{eq:group_sequential_bound} \end{equation} 
\end{theorem}


Theorem \ref{theorem:gst_bound} identifies GST as a superior middle ground. By optimizing over $K$ groups rather than $M$ datasets, GST suppresses the impact of inter-group heterogeneity by factors of $1/K^{1/3}$ and $1/K$. Crucially, as shown in Lemma \ref{lemma:group_heterogeneity}, proper grouping ensures that the inter-group divergence $(\beta_g, \zeta_g)$ remains a controlled fraction of the global heterogeneity. Meanwhile, the joint training within each group retains the variance reduction benefit ($\sigma_k < \sigma$) typical of parallel mixing. Consequently, the GST bound is strictly tighter than that of fully parallel training in Eqn.~\eqref{eq:parallel_bound}, which is hindered by large global heterogeneity, and fully sequential training in Eqn.~\eqref{eq:sequential_bound}, which suffers from unmitigated stochastic noise.

\subsection{Affinity-Aware Grouped Training}

To implement the GST strategy, we require an affinity-aware mechanism to partition datasets. Following our convergence analysis, our primary objective is to cluster datasets with high optimization affinities to minimize the intra-group divergence $(\beta_k, \zeta_k)$. This ensures that the parallel mixing within each group remains stable, while the residual inter-group heterogeneity is effectively compressed through the sequential stages of GST.

\paragraph{Gradient-Based Affinity.} We quantify affinity via the pairwise distance between dataset-level gradients. Based directly on the notion of heterogeneity in Eqn.~\eqref{eqn:gvariance}, we define the dataset distance as:
\begin{align}
    Dist(T_m,T_n)=\|\nabla F_m(\Theta) - \nabla F_n(\Theta)\|^2,
    \label{eq:gradient}
\end{align}
where $\sum_{m,n=1}^M 
\|\nabla F_m(\Theta) - \nabla F_n(\Theta)\|^2=M\cdot\sum_{m=1}^M\ 
\|\nabla F_m(\Theta) - \nabla F(\Theta)\|^2$. 

A smaller distance signifies a higher affinity, indicating that the datasets share consistent optimization trajectories. In practice, we estimate $Dist(D_m, D_n)$ through a controlled probing procedure: starting from a common initialization, we finetune the model on each dataset for a fixed number of steps and accumulate the gradients. We then apply spectral clustering on the resulting distance matrix to form the task groups $\{\mathcal{G}_k\}$.


\paragraph{Alternative: Empirical Transferability.} We also consider an empirical transferability based criterion \cite{zamir2018taskonomy} as a complementary baseline.\footnote{A discussion of the affinity metrics and dataset relationships is included in Appendix~\ref{app:relation}.} This measure evaluates affinity by observing cross-dataset performance gains: 
\begin{align}
    &Trf(D_m\to D_n) = \frac{F_n(\Theta_m)-F_n(\Theta_n)}{F_n(\Theta_n)}.
    \label{eq:taskonomy}
\end{align}
Here $\Theta_m$, $\Theta_n$ denote the model parameters after training on $D_m$ and $D_n$ respectively. While intuitive, this metric is computationally expensive (requiring $M^2$ testing pairs) and less directly coupled with the optimization dynamics in Theorem~\ref{theorem:gst_bound}. We thus primarily use it for comparative analysis. 

\paragraph{Training Implementation: From Sequential to Progressive.} While Theorem~\ref{theorem:gst_bound} assumes strict sequential transitions, repeatedly cycling through groups after short update intervals can present significant practical hurdles to large-scale ALLM training, entailing catastrophic forgetting and prohibitive data loading overhead. To bridge this gap, we adopt in practice a Progressive Grouped Training schedule: we begin with group $\mathcal{G}_1$ and incrementally expand the training pool to $\bigcup_{j=1}^{k+1} \mathcal{G}_j$ at each subsequent stage. This progressive expansion serves as an efficient approximation of GST; it mitigates initial gradient conflicts through sequential introduction while maintaining the stability and I/O throughput of joint mixing. Consequently, this hybrid regime balances the variance reduction of parallel training with the heterogeneity suppression of the sequential regime.

\section{Experiments}

\begin{table*}[htb]
\setlength{\tabcolsep}{3pt}
    \renewcommand{\arraystretch}{1.1}
    \centering
\resizebox{1.0\textwidth}{!}{%
\begin{tabular}{c|ccccccccccccccccc}
\toprule

 & \multicolumn{1}{c}{D1} & \multicolumn{1}{c}{D2} & \multicolumn{1}{c}{D3} & \multicolumn{1}{c}{D4} & \multicolumn{1}{c}{D5} & \multicolumn{1}{c}{D6} & \multicolumn{1}{c}{D7} & \multicolumn{1}{c}{D8} & \multicolumn{1}{c}{D9} & \multicolumn{1}{c}{D10} & \multicolumn{1}{c}{D11} & \multicolumn{1}{c}{D12} & \multicolumn{1}{c}{D13} & \multicolumn{1}{c}{D14} & \multicolumn{1}{c}{Avg} & \multicolumn{1}{c}{W.Avg} & \multicolumn{1}{c}{$\mathcal{T.E.}$} \\ \midrule
SALMONN & 50.7 & 39.4 & 52.4 & 88.9 & 90.8 & 55.1 & 35.4 & 70.0 & 66.3 & 61.4 & 9.9 & 40.5 & 39.8 & 53.8 & 53.9 & 54.1 & N/A\\
Individual & 60.1 & 96.7 & 47.5 & 98.0 & 91.3 & 96.5 & 44.6 & 92.8 & 62.9 & 90.8 & 77.5 & 64.9 & 83.6 & 63.1 & 76.4 & 77.0 & \textasciitilde7d\\
Sequential & 42.7 & 35.2 & 47.0 & 61.6 & 92.4 & 28.4 & 33.7 & 56.0 & 56.9 & 63.5 & 14.6 & 45.6 & 41.0 & 61.1 & 48.6 & 54.3 &\textasciitilde7d\\
Mix All & 56.1 & 94.9 & 50.6 & 97.6 & 87.9 & 95.0 & 47.4 & 89.5 & 67.1 & 87.6 & 72.8 & 55.4 & 76.5 & 62.2 & 74.3 & 74.2 &\textasciitilde4d\\
GST-T2 & 58.8 & 96.2 & 52.8 & 97.9 & 89.0 & 95.9 & 48.4 & 90.6 & 66.3 & 90.9 & 73.3 & 60.6 & 72.9 & 62.2 & 75.4 & 74.5 &\textasciitilde2d\\
GST-G3 & 56.7 & 96.0 & 52.4 & 98.0 & 90.1 & 95.8 & 48.6 & 88.8 & 66.2 & 88.2 & 73.7 & 57.3 & 79.6 & 62.1 & 75.2 & 75.0 &\textasciitilde2d\\
GST-G2 & 55.4 & 95.1 & 51.9 & 97.6 & 88.3 & 95.5 & 47.4 & 91.0 & 67.4 & 87.2 & 70.8 & 57.0 & 79.9 & 61.6 & 74.7 & 74.6 &\textasciitilde2d\\ \midrule
Mix All       & 52.1 & 93.1 & 46.3 & 93.9 & 89.5 & 92.1 & 41.4 & 91.4 & 60.5 & 81.2 & 67.7 & 45.8 & 54.1 & 52.9 & 68.7 & 63.9  & \textasciitilde0.5d                                    \\
GST-T2          & 50.9 & 94.7 & 46.9 & 96.0 & 87.1 & 93.2 & 41.1 & 91.6 & 61.4 & 82.9 & 69.4 & 43.8 & 53.8 & 53.8 & 69.0 & 64.7  & \textasciitilde0.5d                                     \\
GST-G3        & 48.2 & 95.2 & 46.5 & 96.4 & 88.9 & 93.7 & 41.2 & 91.6 & 58.9 & 81.5 & 69.0 & 52.5 & 51.1 & 52.1 & 69.1 & 63.4  &  \textasciitilde1d                                    \\
GST-G2         & 50.4 & 93.6 & 46.1 & 96.1 & 89.3 & 93.4 & 40.1 & 92.0 & 60.6 & 82.0 & 69.5 & 44.1 & 53.6 & 51.0 & 68.7 & 63.5  & \textasciitilde0.5d                                     \\
\bottomrule

\end{tabular}%
}
  \caption{\textbf{Token-Level Accuracy(\%) Comparison in Full-Data Training (Up) and Low-Resource Finetuning (Down).} $D_1 \dots D_{14}$ correspond to: AudioCaps, ChimeHome, Clotho, CochlScene, IEMOCAP, Jamendo, MACS, MusicNet, MusicQA, OpenAQA, PromptSpeech, SoundDescs, TextrolSpeech, and WavCaps, respectively.}
  \label{tab:comparison}
\end{table*}

\begin{figure*}[htbp]
  \centering
  \begin{minipage}{0.22\textwidth}
    \includegraphics[width=\textwidth]{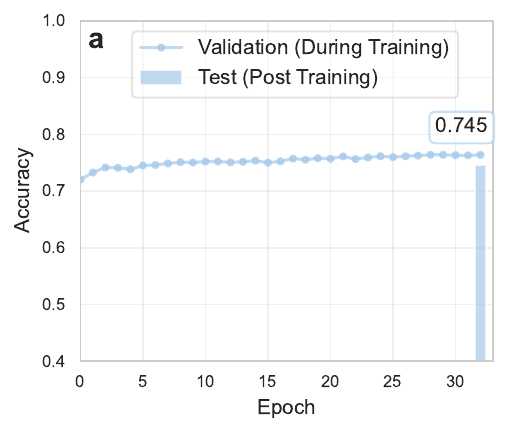}
  \end{minipage}
  \begin{minipage}{0.25\textwidth}
    \includegraphics[width=\textwidth]{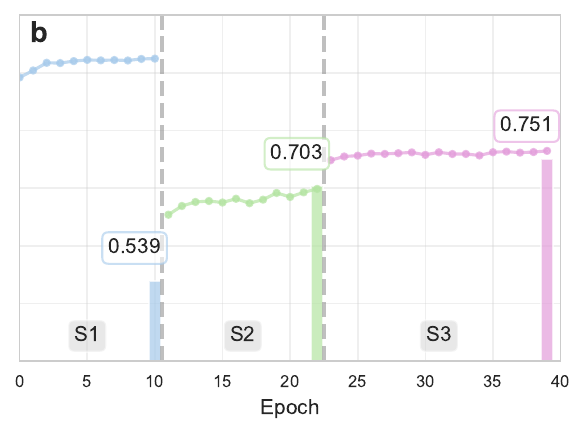}
  \end{minipage}
  \begin{minipage}{0.51\textwidth}
    \includegraphics[width=\textwidth]{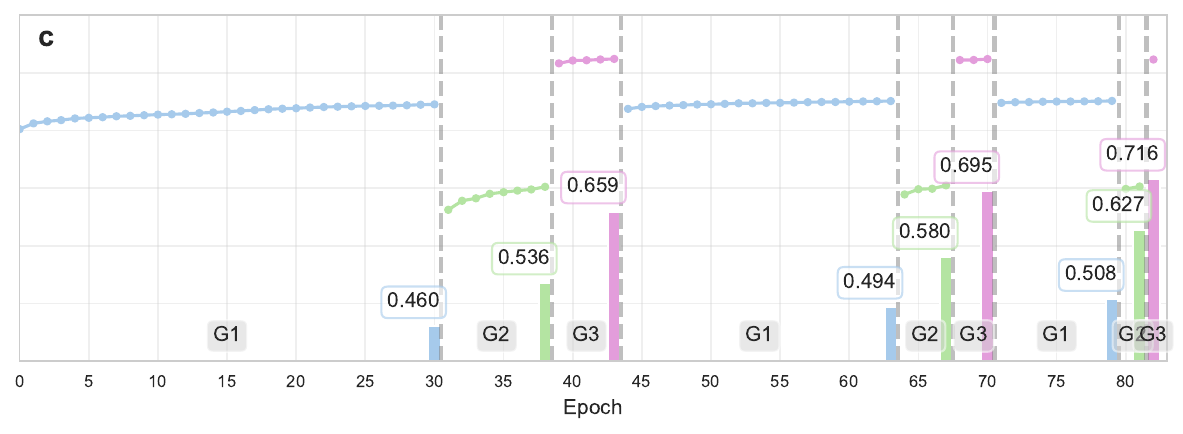}
  \end{minipage}
  \vspace{-0.3cm}
  \caption{\textbf{Comparison of training dynamics and convergence across different scheduling protocols.} (a), (b), and (c) represent Mix-all, GST (Progressive), and GST (Strict Sequential), respectively. The curves denote the validation accuracy across training epochs, while the bars indicate the average test accuracy across all datasets at the end of each stage.}
  \label{fig:convergence}
\end{figure*}

\subsection{Experimental Setup}

We empirically evaluate our proposed data-level training strategy to verify if the theoretical advantages of GST translate into superior performance and convergence in ALLM training. 

\paragraph{Model Architecture.} We select SALMONN-13B \cite{tang2024salmonn} as our experimental platform. While newer models have emerged, SALMONN remains a highly representative and open-source foundation for ALLMs, featuring a standard architecture that combines a Whisper encoder with a Vicuna-13B backbone. Its well-documented training pipeline makes it an ideal testbed for isolating the effects of dataset scheduling from architectural innovations. We strictly follow the official configuration to ensure all gains stem solely from GST strategy. The performance is then evaluated by token-level prediction accuracy.

\paragraph{AudioQA Dataset Formulation.} We evaluate our method on 14 diverse datasets covering speech, music, and environmental sounds. Following \cite{openAQA}, all tasks are unified into an AudioQA format. This collection serves as a benchmark for structured heterogeneity in general audio perception, with significant variations in annotation style and domain-specific knowledge. Dataset descriptions are detailed in Appendix~\ref{app:dataset}.

\paragraph{Training Strategies.} We evaluate several canonical dataset scheduling strategies. To ensure a rigorous controlled experiment, all methods share identical model architectures, optimization settings, and training budgets, differing only in their dataset organization:\footnote{Training details are provided in Appendix~\ref{app:training}.} \textbf{Parallel Training:} Jointly optimizes by uniformly sampling data from all 14 datasets at each iteration; \textbf{Sequential Training:} Optimizes datasets one at a time in a fixed order, representing a zero-mixing extreme; \textbf{Independent Training:} Finetunes a separate model for each dataset, serving as an empirical upper bound for single-task performance; \textbf{GST (Proposed)}: Our affinity-aware strategy. To explore the granularity and criteria of grouping, we evaluate three variants, i.e. \textbf{GST-T2}, Taskonomy-based grouping ($K=2$) using empirical transferability (Eqn.~\ref{eq:taskonomy}), and \textbf{GST-G2 / G3}, Gradient-based grouping ($K=2$ / $K=3$) using gradient distance (Eqn.~\ref{eq:gradient}). Unless otherwise noted, we employ the Progressive implementation for stability.





\begin{table}[htbp]
\setlength{\tabcolsep}{3pt}
    \renewcommand{\arraystretch}{1.1}
\centering
\resizebox{0.45\textwidth}{!}{%
\begin{tabular}{c|cc|cc}
\toprule
 Schedule & Avg    & W.Avg    & Avg    & W.Avg    \\
  \midrule
Progressive* & 75.2 & 75.0 & 74.7 & 74.6 \\
Reverse Progressive & 72.1 & 72.1 & 72.9 & 72.6 \\
Strict Cycle Sequential & 70.7 & 64.0 & 60.3 & 68.2 \\
\bottomrule
\end{tabular}
}
\caption{\textbf{Accuracy(\%) Comparison of Ordering Strategies in Full-Data Training.} We evaluate different curriculums using the grouping criteria of GST-G3 (left) and GST-G2 (right).``*" denotes our method.}
  \label{tab:ordering}
  \vspace{-0.5cm}
\end{table}

\subsection{Main Results}

We evaluate training strategies under two distinct settings to comprehensively assess their impact on optimization:

\noindent\textbf{Full-Data Training:} Utilizing all available data and training from scratch, this setting emulates the large-scale joint training phase typical in building foundation models.

\noindent\textbf{Low-Resource Finetuning:} Capping each dataset at $N=250$ samples and initializing from SALMONN checkpoints, this setting emulates  supervised finetuning of models with scarce data.

Tab.~\ref{tab:comparison} summarizes the performance across 14 AudioQA datasets. We report the raw testing accuracy for each task, along with the cumulative time elapsed ($\mathcal{T.E.}$) and two aggregate metrics: the unweighted average accuracy (Avg), and the sample-size weighted average accuracy (W.Avg).

\subsubsection{Performance on Full-Data Training}

As illustrated in Tab.~\ref{tab:comparison}, the performance of GST strategies in full-data training is characterized by:

\noindent\textbf{Superior Performance.} Our proposed GST variants consistently outperform Parallel Training (Mix-all). Specifically, GST-T2 achieves the highest unweighted average, while GST-G3 leads in the weighted average. Notably, all variants significantly exceed the Original SALMONN baseline and closely approach the Independent Training reference (the empirical upper bound), underscoring the validity of our experimental comparison.

\noindent\textbf{Robustness Against Forgetting.} Compared to Sequential Training, which suffers from a drastic drop in accuracy due to catastrophic forgetting, our GST implementation maintains a high performance ceiling, confirming that the progressive introduction of groups serves as an effective buffer.

\noindent\textbf{Enhanced Optimization Efficiency.} A key highlight of our results is the significant reduction in training time. While the parallel baseline requires approximately 4 days to reach convergence, GST variants complete the process within 2--3 days. This 30--40\% speedup in large-scale training empirically validates the tighter convergence bounds derived in Theorem~\ref{theorem:gst_bound}.

\subsubsection{Stability in Low-Resource Scenarios}

In the low-resource regime, the primary challenge shifts from managing massive heterogeneity to ensuring sample efficiency. We omit the Sequential baseline here as it leads to severe forgetting with sparse data, yielding non-informative results. Findings for the remaining strategies are as follows:

\noindent\textbf{Performance Stability.} In the low-resource regime, GST variants achieve comparable performance to the Parallel baseline. This result notably demonstrates that the group-wise sequential nature of GST does not compromise the model's adaptation capability even under extreme data sparsity.

\noindent\textbf{Diminishing Acceleration Returns.} Unlike the full-data regime, the training acceleration effect is less pronounced here, with marginal reductions in wall-clock time. This phenomenon can be attributed to a trade-off between heterogeneity suppression and early-stage overfitting: when training on a small initial group $\mathcal{G}_1$, the model may rapidly overfit to these sparse samples before the data pool expands. Consequently, the potential time gains from reduced gradient variance are partially offset by the optimization overhead required to stabilize learning on such limited data.

\subsection{Discussion}

\subsubsection{Convergence Behaviors}

To further investigate how GST features in optimization dynamics, we plot  in Fig.~\ref{fig:convergence} the training of three representative strategies: Mix-all, GST-G3 (Strict Sequential), and GST-G3 (Progressive).
As depicted in the curves, Strict Sequential training exhibits a ``volatile" pattern; while performance peaks at the end of each stage, it suffers drastic drops upon returning to previous data groups (e.g., from Stage 3 back to Stage 1), characterizing classic catastrophic forgetting. This instability forces the model to undergo significantly more training cycles to reach a stable state. In contrast, Progressive GST maintains a steady and monotonic improvement in test accuracy, proving the effectiveness of our progressive implementation. Compared with Mix-All, Progressive GST also reaches a stable performance plateau slightly earlier (approx. epoch 25 vs. epoch 27). More importantly, the per-epoch training time for Mix-all is substantially longer than that of GST. We attribute this to the high gradient heterogeneity in Mix-all, which forces the optimizer to reconcile conflicting update directions within every batch, leading to higher computational overhead and slower stable descent.

\subsubsection{Impact of Group Ordering: The Stability-First Curriculum}

To examine how the temporal sequence of dataset groups influences the optimization trajectory, we compare three scheduling strategies: Strict Cycle Sequential, Progressive, and Reverse Progressive (see Tab.~\ref{tab:ordering}). Our analysis reveals that the effectiveness of GST is highly sensitive to the initial training phase and the data retention mechanism. The superiority of the positive Progressive order over the Reverse variant underscores a "stability-first" curriculum effect. When high-variance or outlier tasks are introduced prematurely (as in the Reverse order), the model encounters a "noisy" initialization phase that is both computationally expensive and optimizationally destructive. A complex initial group not only requires excessive epochs to reach marginal convergence but also forces the model into suboptimal regions of the loss landscape. Furthermore, the poor performance of Strict Cycle Sequential highlights the danger of abrupt distribution shifts introduced by the original sequential strategy. These results confirm that an affinity-aware initialization and optimization stability are essential for navigating the complex heterogeneity of multi-dataset ALLM training.

\section{Conclusion}

In this paper, we introduced Grouped Sequential Training (GST), a principled optimization framework designed to mitigate the challenges of gradient heterogeneity in multi-dataset ALLM training. By strategically grouping datasets based on affinity measures and introducing them via a progressive scheduling protocol, GST effectively balances inter-dataset interference and optimization efficiency.
Theoretically, we derived tighter convergence bounds for GST, demonstrating that reducing intra-group gradient variance directly accelerates stable descent. Empirically, evaluations across 14 diverse AudioQA datasets show that GST variants not only outperform parallel and sequential training baselines but also achieve a 30--40\% reduction in training time. Our findings underscore the importance of affinity-aware scheduling in large-scale joint training and provide a scalable, interpretable pathway for developing robust ALLMs.

\section{Limitations}

While the proposed GST framework demonstrates significant acceleration and stability in ALLM training, we acknowledge several limitations that open avenues for future research.

\noindent\textbf{Diversity of Backbones and Frameworks.} Our current evaluation is primarily conducted using the SALMONN framework with a Vicuna-13B backbone. Although these are representative of state-of-the-art audio-language models, the scalability of GST to even larger models (e.g., 70B parameters) or different architectural paradigms (e.g., MoE-based audio LLMs) remains to be extensively verified. Future work will investigate whether the gradient affinity patterns observed here persist across diverse model scales and initialization states.

\noindent\textbf{Bottleneck Analyses in Mix-all Training Efficiency.} We observed that the Mix-all baseline exhibits slower per-epoch training speeds compared to partitioned strategies. Our profiling indicates this is not merely due to data loading overhead, but stems from gradient variance explosion in the early stages of training. When heterogeneous datasets are shuffled into a single stream, the frequent directional shifts in gradients ($\nabla F_m$) lead to unstable optimizer states and more rejected updates in the loss landscape. GST mitigates this by grouping consistent signals, but a more rigorous theoretical quantification of this "mixing penalty" is still needed.

\noindent\textbf{Refinement of the "Stability-First" Ordering.} Our Stability-First ordering strategy (beginning with high-affinity, low-variance groups) is based on the heuristic that establishing a robust representation early on prevents computational waste and optimizational destruction. However, a purely optimal ordering might require a more dynamic approach. The current static ordering, while effective, does not account for the temporal evolution of dataset relationships—as the model learns, two datasets that were initially conflicting might become synergistic. Exploring a dynamic, feedback-driven ordering mechanism is a promising direction.
\bibliography{custom}
\newpage
\appendix

\section{Appendix}
\label{sec:appendix}

\subsection{Theoretical Proof}
\subsubsection{Proof of Prop.~\ref{prop:decomposition} (Heterogeneity Decomposition)}

For any dataset $m$ belonging to group $\mathcal{G}_k$, we decompose the deviation from the global mean by inserting the group mean $\nabla F^{(k)}$:
\begin{align} &\|\nabla F_m - \nabla F\|^2 \\&= \|(\nabla F_m - \nabla F^{(k)}) + (\nabla F^{(k)} - \nabla F)\|^2 \end{align} Expanding the squared norm: 
\begin{align} &\|\nabla F_m - \nabla F\|^2\\ &= \|\nabla F_m - \nabla F^{(k)}\|^2 + \|\nabla F^{(k)} - \nabla F\|^2 \\&+ 2\langle \nabla F_m - \nabla F^{(k)}, \nabla F^{(k)} - \nabla F \rangle \end{align}
Summing over all datasets $m \in \{1, \dots, M\}$:
\begin{align}
    &\sum_{M=1}^m \|\nabla F_m - \nabla F\|^2 \\
    &=\sum_{k=1}^K \sum_{m \in \mathcal{G}k} \|\nabla F_m - \nabla F\|^2 \\&= \sum_{k=1}^K \sum_{m \in \mathcal{G}k} \|\nabla F_m - \nabla F^{(k)}\|^2 \\&+ \sum_{k=1}^K \sum_{m \in \mathcal{G}k} \|\nabla F^{(k)} - \nabla F\|^2 + \text{Cross Terms} \\
    &=\sum_{k=1}^K \sum_{m \in \mathcal{G}k} \|\nabla F_m - \nabla F^{(k)}\|^2 \\&+ M_k\sum_{k=1}^K \|\nabla F^{(k)} - \nabla F\|^2+ \text{Cross Terms},
\end{align}
where $\text{Cross Terms}=0$, due to:
\begin{align} &2 \sum_{m \in \mathcal{G}k} \langle \nabla F_m - \nabla F^{(k)}, \nabla F^{(k)} - \nabla F \rangle \\&= 2 \langle \underbrace{\sum_{m \in \mathcal{G}k} (\nabla F_m - \nabla F^{(k)})}_{\mathbf{0}}, \nabla F^{(k)} - \nabla F \rangle = 0. \end{align}
Therefore, we have
\begin{align}
\small
&\frac{1}{M} \sum_{m=1}^M \|\nabla F_m - \nabla F\|^2 \\ = &\frac{1}{M} \sum_{k=1}^K \sum_{m \in \mathcal{G}_k} \|\nabla F_m - \nabla F^{(k)}\|^2 \\+& \frac{1}{M} \sum_{k=1}^K M_k \|\nabla F^{(k)} - \nabla F\|^2.
\end{align}

\subsubsection{Proof of Lemma~\ref{lemma:group_heterogeneity} (Heterogeneity Bound)}
\label{proof:lemma2}
For intra-group heterogeneity, we have:
\begin{align}
&\frac{1}{K}\sum_{k=1}^K\|\nabla F^{(k)}(\Theta)-\nabla F(\Theta)\|^2 \nonumber\\
    &\le\frac{M}{KM_{min}}\cdot\frac{1}{M}\sum_{k=1}^KM_k\|\nabla F^{(k)}(\Theta)-\nabla F(\Theta)\|^2 \nonumber \\
    &\le \frac{M\beta^2}{KM_{min}} \|\nabla F(\Theta)\|^2 + \frac{M\zeta^2}{KM_{min}} \nonumber\\ 
    &- \frac{1}{KM_{min}}\sum_{k=1}^K\sum_{m \in \mathcal{G}_k}\|\nabla F_m(\Theta) - \nabla F^{(k)}(\Theta)\|^2.
    \label{eqn:variance_decompose}
\end{align}
Comparing Eqn.~\eqref{eqn:variance_decompose} with the form of bound in Eqn.~\eqref{eqn:groupvariance}, we have $\zeta_g^2<\frac{M\zeta^2}{KM_{min}}- \Gamma_{\mathcal{G}}$ and $\beta_g^2<\frac{M\beta^2}{KM_{min}}-\Delta_{\mathcal{G}}$, with $M_{min}$ denoting the smallest group size and $\Gamma_{\mathcal{G}}, \Delta_{\mathcal{G}} \ge 0$ are the residual terms representing the heterogeneity absorbed within groups via affinity-based alignment.

\subsubsection{Proof of Stochastic Gradient Variance}
\label{proof:variance}

Noting the stochastic gradient on dataset m as $g_m(\Theta)$ and stochastic gradient in $\mathcal{G}_k$ as $g^{(k)}(\Theta)$, for the intra-group stochastic gradient variance, we have:
\begin{align}
    &\|g^{(k)}-\nabla F^{(k)}\|^2\\
    &=\|\frac{1}{M_k}\sum_{m\in\mathcal{G}_k}(g_m-\nabla F_m) \\&+ \frac{1}{M_k}\sum_{m\in\mathcal{G}_k} (\nabla F_m-\nabla F^{(k)})\|^2\\
    &\le \|\frac{1}{M_k}\sum_{m\in\mathcal{G}_k}(g_m-\nabla F_m)\|^2\\& + \|\frac{1}{M_k}\sum_{m\in\mathcal{G}_k} (\nabla F_m-\nabla F^{(k)})\|^2,
\end{align}
with
\begin{align}
    &\mathbb{E}\|\frac{1}{M_k}\sum_{m\in\mathcal{G}_k}(g_m-\nabla F_m)\|^2\\
    &\le\frac{1}{M_k^2} \sum_{m\in\mathcal{G}_k}\mathbb{E}\|g_m-\nabla F_m\|^2 \le \frac{\sigma^2}{M_k},\\
    &\|\frac{1}{M_k}\sum_{m\in\mathcal{G}_k} (\nabla F_m-\nabla F^{(k)})\|^2\\
    &\le \frac{1}{M_k}\sum_{m\in\mathcal{G}_k}\|\nabla F_m-\nabla F^{(k)}\|^2\\
    & \le \beta_k^2\|\nabla F^{(k)}\|^2+\zeta_k^2.
\end{align}
Hence, we have 
\begin{align}
    &\mathbb{E}\left[\|g^{(k)}-\nabla F^{(k)}\|^2\right] \\
    &\le \frac{\sigma^2}{M_k}+\beta_k^2\|\nabla F^{(k)}\|^2+\zeta_k^2=\sigma_k^2.
\end{align}
Here $\sigma_k$ is given by $\frac{\sigma^2}{M_k}$ and intra-group heterogeneity $(\beta_k,\zeta_k)$, and $\sigma_k<\sigma$ can be ensured with bounded intra-group heterogeneity.

\subsection{Dataset Description}
\label{app:dataset}

To evaluate the proposed GST framework across diverse acoustic scenarios, we curate a comprehensive benchmark consisting of 14 datasets covering ambient sounds, speech, and music. These datasets are categorized into three primary types of tasks: Audio Captioning (CAP), Audio Question Answering (AQA), and Audio Classification (CLS). For classification tasks, we follow a standard ``Classification to QA" prompting strategy to ensure compatibility with LLM instruction-following. For classification tasks, we follow a standard ``Classification to QA" prompting strategy to ensure compatibility with LLM instruction-following. Specifically, we list the datasets below:
\begin{itemize}
    \item \textbf{Ambient \& General Audio:} We include AudioCaps \cite{kim2019audiocaps}, Clotho\_v2 \cite{drossos2020clotho}, WavCaps \cite{mei2024wavcaps}, and MACS \cite{martin2021ground_macs} for general audio captioning. SoundDescs \cite{koepke2022sounddescs} provides fine-grained descriptions, while OpenAQA \cite{openAQA} (derived from AudioSet \cite{gemmeke2017audioset}) offers large-scale temporal audio reasoning.
    \item \textbf{Acoustic Scenes:} ChimeHome \cite{foster2015chime} and CochlScene \cite{jeong2022cochlscene} are used for scene recognition. To adapt these CLS datasets for ALLM training, we hard-code labels into natural language QA pairs (e.g., "Where is the audio recorded?" $\rightarrow$ "The audio is recorded in a car.").
    \item \textbf{Music Analysis:} MusicQA \cite{liu2024musicqa}, MusicNet \cite{thickstun2016learning_musicnet}, and Jamendo \cite{koh2025jamendo} cover music captioning, ensemble recognition, and mood analysis.
    \item \textbf{Speech \& Emotion}: IEMOCAP \cite{busso2008iemocap} is utilized for emotion recognition using SALMONN annotations. PromptSpeech \cite{guo2023promptspeech} and TextrolSpeech \cite{ji2024textrolspeech} provide speech-to-captioning tasks focusing on speaker traits and style.
\end{itemize}

\begin{table*}[ht]
\centering
\caption{Statistical overview of the multi-dataset benchmark used in this study. $D_1$ to $D_{14}$ cover a diverse range of audio domains and tasks.}
\label{tab:dataset_stats}
\small
\resizebox{1.0\textwidth}{!}{%
\begin{tabular}{c|ccccc}
\toprule
\textbf{Dataset} & \textbf{Task Type} & \textbf{Primary Domain} & \textbf{Train} & \textbf{Val} & \textbf{Test} \\
\midrule
AudioCaps & CAP & General Audio & 30,741 & 3,804 & 3,804 \\
ChimeHome & CLS $\rightarrow$ QA & Household & 4,378 & 879 & 880 \\
Clotho\_v2 & CAP & General Audio & 3,839 & 1,045 & 1,045 \\
CochlScene & CLS $\rightarrow$ QA & Urban Scenes & 60,855 & 7,573 & 7,687 \\
IEMOCAP & CLS $\rightarrow$ QA & Speech Emotion & 3,272 & 409 & 409 \\
Jamendo & QA & Music & 26,090 & 3,261 & 3,261 \\
MACS & CAP & Ambient Sound & 3,144 & 393 & 393 \\
MusicNet & QA & Classical Music & 264 & 33 & 33 \\
MusicQA & QA & General Music & 42,867$^*$ & 70,011 & 5,040 \\
OpenAQA & Temporal QA & Multi-domain & 213,158 & 26,644 & 26,644 \\
PromptSpeech & CAP & Speech Style & 21,272 & 2,658 & 2,658 \\
SoundDescs & CAP & Environmental & 26,452 & 3,306 & 3,306 \\
TextrolSpeech & CAP & Speech Style & 188,717 & 23,589 & 23,589 \\
WavCaps & CAP & Large-scale Web & 321,420 & 40,177 & 40,178 \\
\midrule
\textbf{Total} & - & - & \textbf{946,469} & \textbf{183,782} & \textbf{119,030} \\
\bottomrule
\end{tabular}}
\vspace{2pt}
\flushleft{\footnotesize $^*$ For MusicQA, we use the pretrain (42,867) for training and finetune (70,011) sets for validation.}
\end{table*}

\begin{table*}[htbp]
\centering
\caption{Comparison of Dataset Relationship Metrics: Taskonomy-based vs. Gradient-based.}
\label{tab:metric_comparison}
\small
\begin{tabular}{c c c}
\toprule
\textbf{Property} & \textbf{Taskonomy-based Metric} & \textbf{Gradient-based Metric (GST)} \\ \midrule
\textbf{Symmetry} & Insymmetric & Inherently symmetric ($A_{i,j} = A_{j,i}$) \\
\textbf{Computational Cost} & Extremely low (Pre-computed) & High (Requires per-dataset backprop) \\
\textbf{Size Sensitivity} & \textbf{Yes}; heavily biased by sample counts & \textbf{No}; focuses on directional alignment \\
\textbf{Stability} & Static throughout the training process & Dynamic; evolves with the model state \\
\textbf{Representativeness} & Semantic and label-level relationships & Optimization and landscape-level essence \\ \bottomrule
\end{tabular}
\end{table*}

\begin{figure*}[htbp]
     \centering
     \begin{subfigure}[b]{0.24\textwidth}
         \centering
         \includegraphics[width=\textwidth]{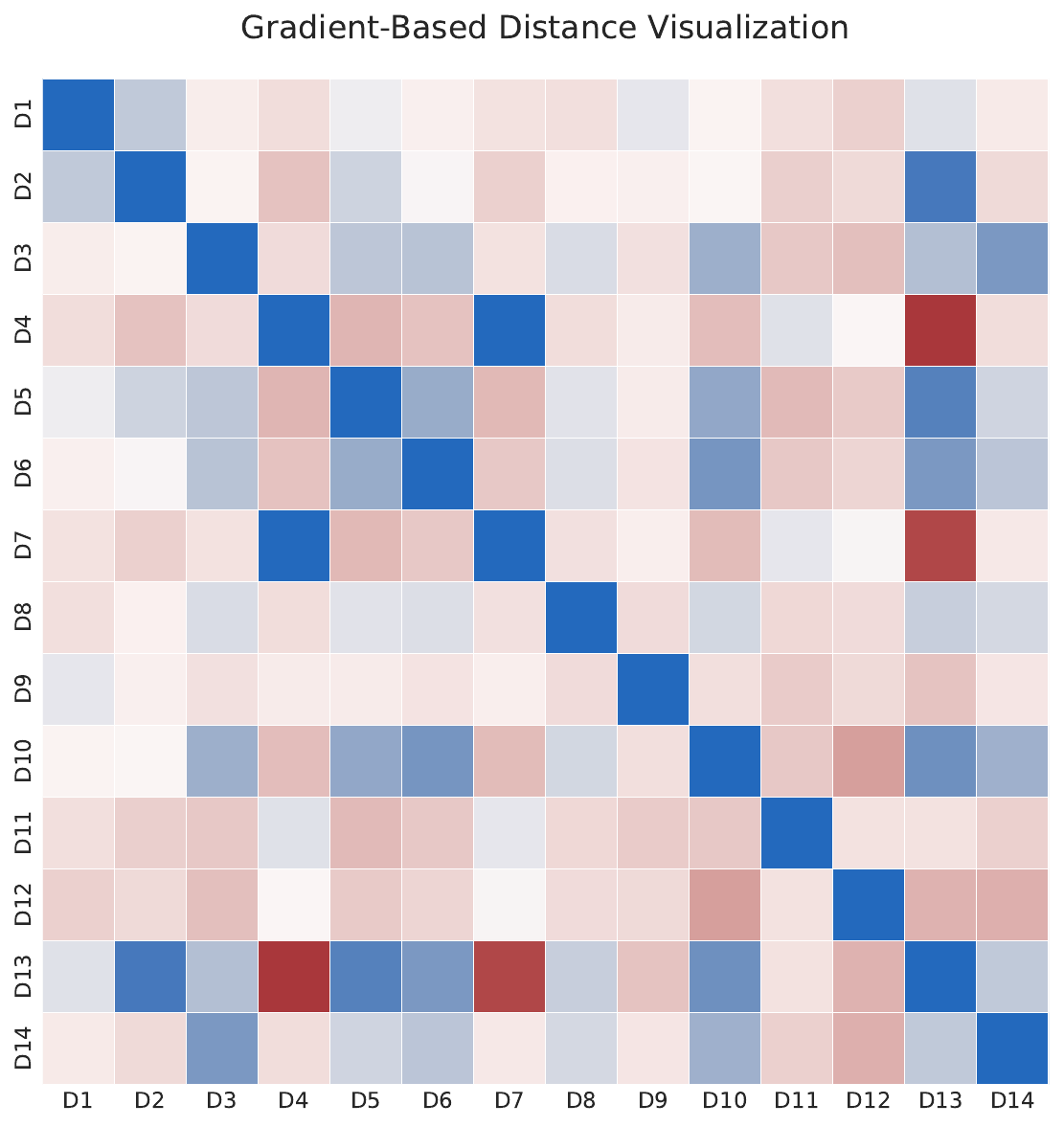}
     \end{subfigure}
     \hfill 
     \begin{subfigure}[b]{0.24\textwidth}
         \centering
         \includegraphics[width=\textwidth]{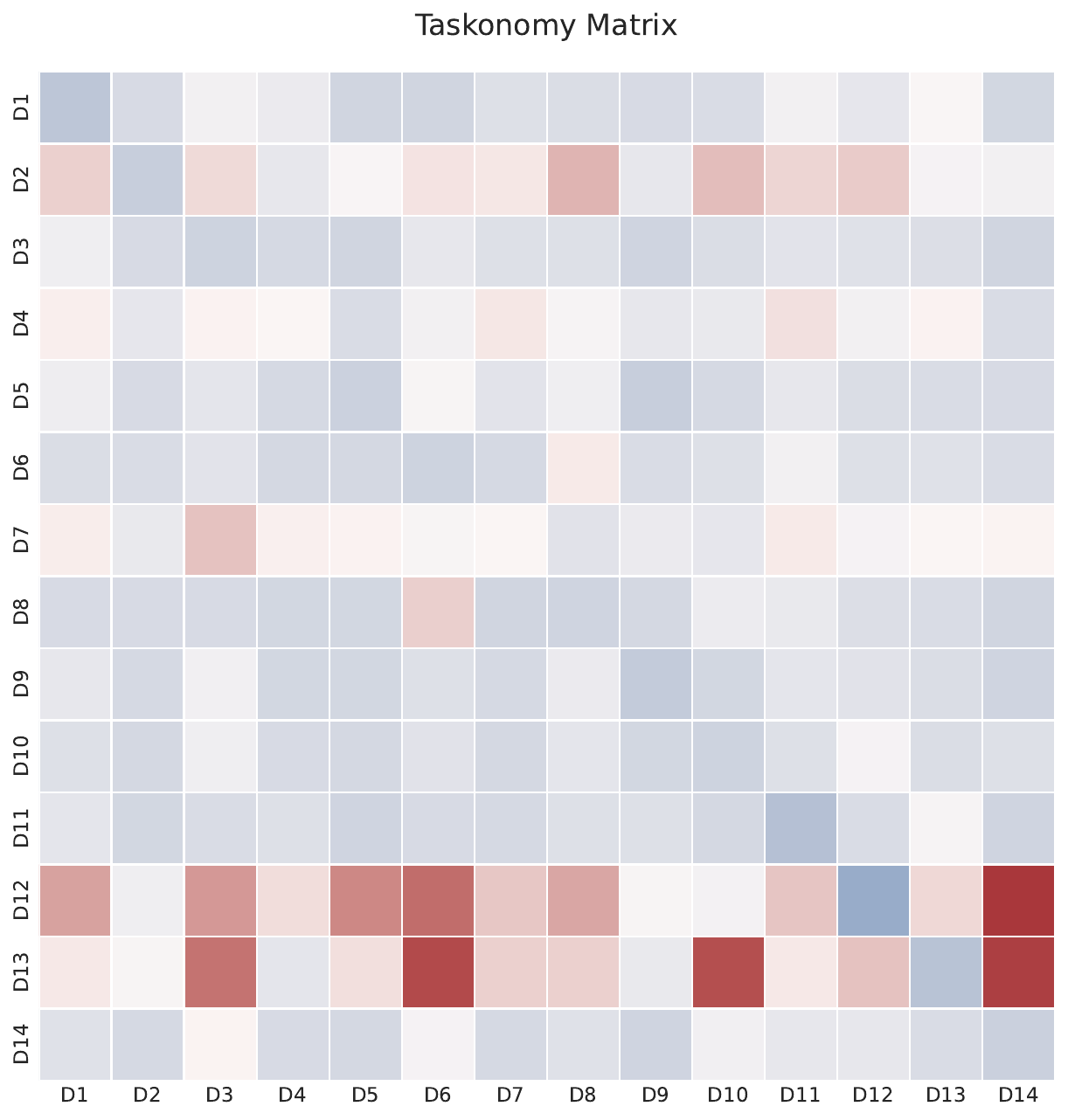}
     \end{subfigure}
     \hfill
     \begin{subfigure}[b]{0.44\textwidth}
         \centering
         \includegraphics[width=\textwidth]{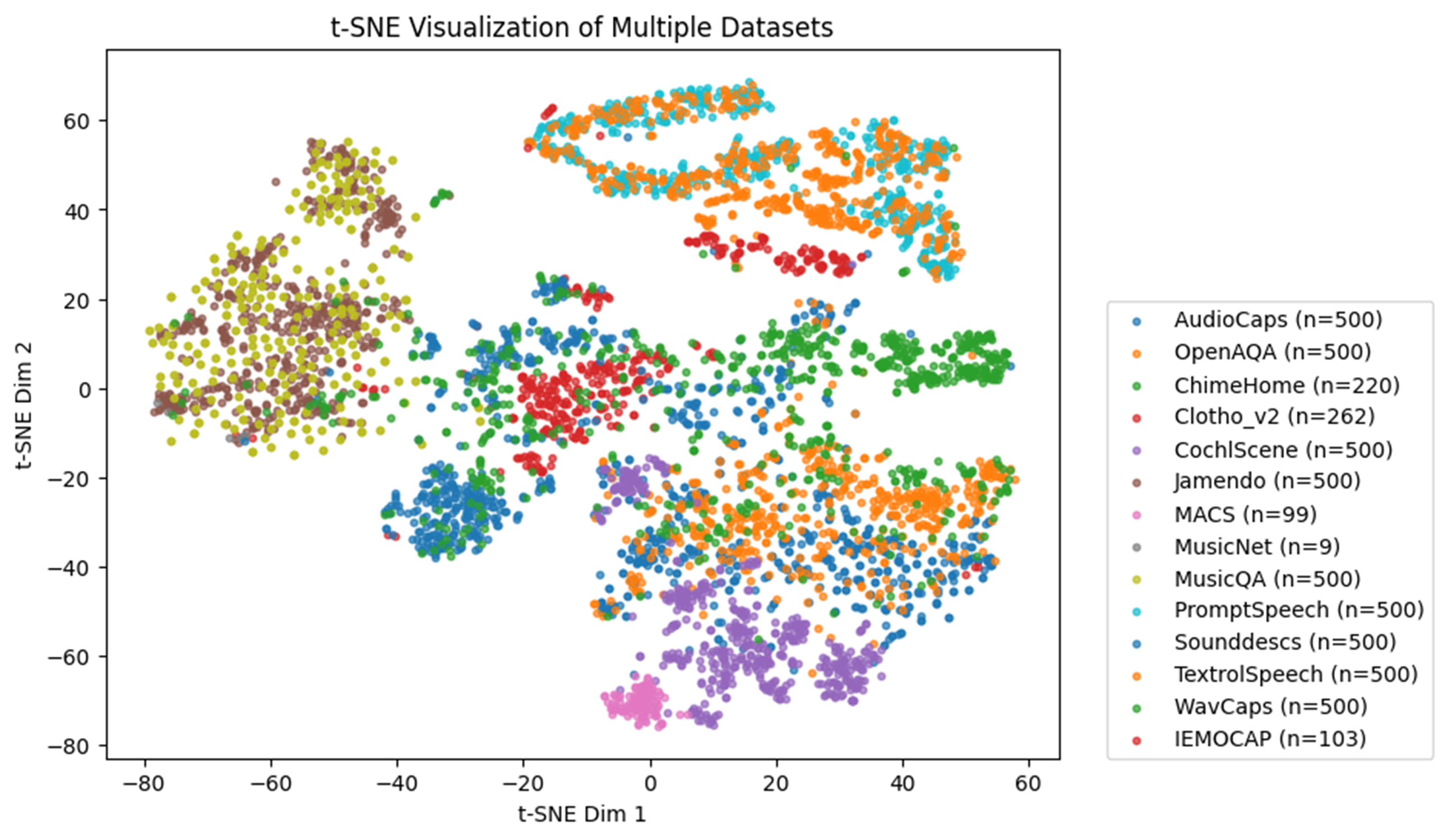}
     \end{subfigure}
     \caption{\textbf{Visualization of Dataset Relationship Measurements.} (a) Taskonomy-based Affinity Matrix and (b) Gradient-based Affinity Matrix, where blue indicates higher similarity (closer relationship) and red indicates lower similarity (further distance). (c) t-SNE Visualization of Acoustic Features, demonstrating distinct clustering based on audio domains (e.g., Music, Speech, Environmental).}
     \label{fig:relation}
\end{figure*}

\subsection{Details of Training Settings}
\label{app:training}

\textbf{Model Architecture \& Configuration.} We build our model based on the SALMONN framework. The architecture consists of a dual-encoder system: a Whisper-Large-v2 for speech feature extraction and a BEATs (iter3+) for audio semantic encoding. Both encoders are kept frozen during training to preserve their pre-trained acoustic representations. A window-level Q-Former is employed to aggregate audio features with a window size and stride of 0.33 seconds, compressing the sequence into a single query token per window.To efficiently adapt the large language model, we utilize Vicuna-13B-v1.1 as the backbone and integrate Parameter-Efficient Fine-Tuning (PEFT) via LoRA. The LoRA adapters are applied with a rank $r=8$ and $\alpha=32$, significantly reducing the number of trainable parameters while maintaining competitive performance.

\textbf{Optimization Details.} The model is optimized using the AdamW optimizer with a weight decay of 0.05 and $\beta_2=0.999$. We employ a learning rate schedule starting with a warmup of 3,000 steps, increasing from 1e-6 to a peak of 3e-5, followed by a cosine decay to a minimum of 1e-5. The training is conducted over a maximum of 30 epochs, with an early stopping patience of 5 epochs based on validation loss to prevent overfitting.

\textbf{Hardware \& Distributed Training.} Our training infrastructure and hyperparameter settings vary based on the data scale:
\begin{itemize}
    \item Full-resource Training: Conducted on a cluster of 4 $\times$ NVIDIA A100 (80GB) GPUs. We use a batch size of 4 per GPU with an accumulation step of 1, resulting in an effective global batch size of 16.
    \item Low-resource Training: For specific ablation studies, we use 2 $\times$ NVIDIA A100 GPUs, with the batch size per GPU and other configurations kept consistent, effectively halving the global batch size to 8.
\end{itemize}
All experiments are implemented using the Singularity container platform for environment consistency. Automatic Mixed Precision (AMP) is enabled to accelerate training and reduce memory footprint.

To evaluate the model's performance and convergence efficiency across the multi-dataset benchmark, we employ Token-level Accuracy as our primary metric. This metric is defined as the ratio of correctly predicted tokens to the total number of target tokens in the sequence: \begin{equation} \text{Acc}_{token} = \frac{\sum \text{Correct Tokens}}{\sum \text{Total Tokens}} \end{equation} Unlike sentence-level metrics which can be overly sensitive to minor variations in decoding, Token-level Accuracy provides a high-resolution view of the model's optimization progress. It directly reflects the model's ability to minimize the cross-entropy loss across heterogeneous datasets, making it an ideal proxy for analyzing the convergence behaviors of different training protocols such as Mix-all and GST.

\subsection{Interpreting Dataset Relationships: Gradient vs. Taskonomy-Based Metrics}
\label{app:relation}

\textbf{Comparison of Metric Characteristics.} As shown in Fig.~\ref{fig:relation}, both metrics capture certain aspects of dataset interdependencies. While showing some coarse consistency, it is notable that Taskonomy is sensitive to dataset scale. For instance, large-scale datasets like WavCaps (321k) and OpenAQA (213k) tend to negatively affect the learning of other datasets, creating "gravity wells" in the affinity matrix. They also differ significantly in their mathematical and computational properties, as shown in Tab.~\ref{tab:metric_comparison}.

\textbf{The Acoustic-Optimization Gap.} To further investigate the "Audio" specificity, we performed a t-SNE visualization on the latent acoustic features extracted from the frozen encoders (Fig.~\ref{fig:relation}).
\begin{itemize}
    \item Acoustic Clustering: The t-SNE plot reveals clear, distinct clusters aligned with human-defined domains: Music (MusicQA, Jamendo), Speech (IEMOCAP, PromptSpeech), and Environmental sounds (ChimeHome, CochlScene).
    \item The Discrepancy: Interestingly, neither our Taskonomy-based nor Gradient-based metrics strictly favor these acoustic domains. A music dataset may show higher gradient affinity with a speech-captioning dataset than with another music-tagging task.
\end{itemize}

This discrepancy indicates that the AudioQA relationship is influenced by factors far more complex than raw acoustic similarity. In a generative ALLM framework, the optimization direction $\nabla F_m$ is a joint product of \textbf{Acoustic spectral features} (the "what" is being heard), \textbf{Instruction complexity and label distribution} (the "how" it is being answered), and \textbf{Backbone alignment state} (the current mapping of audio to Vicuna tokens). While t-SNE captures the "acoustic appearance", our Gradient-based metric captures the "optimization essence".

\end{document}